\documentclass{llncs}
\usepackage[table,xcdraw]{xcolor}
\usepackage[noadjust]{cite}
\usepackage[cmex10]{amsmath}
\usepackage[caption=false,font=footnotesize]{subfig}
\usepackage{url}
\usepackage{graphicx}
\usepackage{amssymb}
\everymath{\displaystyle}
\usepackage{comment}

\usepackage[shortlabels]{enumitem}
\usepackage{nameref}
\usepackage{slashbox}
\usepackage{textcomp}
\usepackage{gensymb}
\usepackage[hypertexnames=false]{hyperref}
\newcommand{\myref}[1]{\hyperref[#1]{\ref*{#1}}}
\usepackage[all]{hypcap}
\usepackage{tikz}
\usepackage{tikz-qtree}
\usetikzlibrary{patterns}
\usetikzlibrary{shapes}
\usepackage{tkz-euclide}
\usepackage{pgfplots}
\pgfplotsset{compat=1.13}
\usepackage{algorithm}
\usepackage{algorithmicx}
\usepackage[noend]{algpseudocode} 
\usepackage{xspace}
\usepackage[outline]{contour}
\contourlength{1.2pt}
\usepackage[normalem]{ulem}
\usepackage{mathtools}
\usepackage{array}
\usepackage{tabu}
\usepackage{lscape}

\newcommand{\DeclareForwardSegment}{\mathrm{FW\_SEG}}
\newcommand{\DeclareBackwardSegment}{\mathrm{BW\_SEG}}
\newcommand{\DeclareForwardArc}{\mathrm{FW\_ARC}}
\newcommand{\DeclareBackwardArc}{\mathrm{BW\_ARC}}
\newcommand{\CheckAndAdjustLastPosition}{ChkAdjPos}
\newcommand{\DeclarePenaltySegment}{\mathrm{P\_SEG}}
\newcommand{\DeclarePenaltyArc}{\mathrm{P\_ARC}}

\algnewcommand{\algorithmicgoto}{\textbf{go to}}
\algnewcommand{\Goto}{\algorithmicgoto\xspace}
\algnewcommand{\Label}{\State\unskip}
\algnewcommand{\Continue}{\textbf{continue}}
\algnewcommand{\MyAnd}{\textbf{and}\xspace}
\algnewcommand{\MyOr}{\textbf{or}\xspace}
\algnewcommand{\Not}{\textbf{not}\xspace}
\algnewcommand{\MyFrom}{\textbf{from}\xspace}
\algnewcommand{\MyTo}{\textbf{to}\xspace}
\algnewcommand{\MyStep}{\textbf{step}\xspace}

\begin{document}

\title{An Efficient Combinatorial Algorithm\\for Optimal Compression of a Polyline\\with Segments and Arcs}
\titlerunning{An Efficient Combinatorial Algorithm for Optimal Compression of a Polyline}

\author{Alexander Gribov}

\institute
{
  Esri\\
  380 New York Street\\
  Redlands, CA 92373\\
  \email{agribov@esri.com}
}

\maketitle

\begin{abstract}
\boldmath
The task of finding the optimal compression of a polyline with straight-line segments and arcs is performed in many applications, such as polyline compression, noise filtering, and feature recognition. Optimal compression algorithms find the best solution using the dynamic programming approach, which requires a significant amount of arc fitting. This paper describes an improvement to the dynamic programming approach by reducing the amount of arc fitting necessary to find the optimal solution. Instead of processing from the second to the last vertices in the dynamic programming approach, the algorithm proceeds forward and skips as many steps as possible without affecting the inference in any way. Such a modification extends the practical application of the algorithm to polylines having arcs with a large number of vertices.
\end{abstract}


\section{Introduction}

Finding the optimal compression of a polyline with straight-line segments and arcs when the resultant polyline is required to be within the specified tolerance from the source polyline has a solution with the worst-case complexity $ O{ \left( N^3 \log{ \left( N \right) } \right) } $, where $ N $ is the number of vertices in the source polyline (see \cite{OptimalCompressionWithArcs, OptimalCompressionOfAPolylineWithSegmentsAndArcs}), which prevents the efficient processing of arcs with many vertices. However, many polylines have a well-defined structure and do not require evaluation of all possible combinations. The simple example would be a noisy polyline with just one arc. Another example with two segments will be explained in Sect.~\myref{section:task}. The algorithm described in this paper is the further development of the algorithm described in \cite{OptimalCompressionOfAPolylineWithSegmentsAndArcs} and will be explained in Sect.~\myref{section:algorithm}. Finding the maximum jump for the next segment or arc and then looking backward from the maximum jump will lead to the algorithm, which skips steps used in the dynamic programming approach. An experimental comparison will be given in Sect.~\myref{section:performance}.

\section{Task\label{section:task}}

\begin{figure} [htb]
  \centering
  \begin{tikzpicture} [scale = 0.3]
    \draw [thick, black]
    (0, 0) --
    (10, -10) --
    (20, 0);

    \draw [blue, fill = blue] ( 0, - 0) circle [radius = 0.15] node [black, below left ] {\small{$ 0 $}};
    \draw [blue, fill = blue] ( 1, - 1) circle [radius = 0.15] node [black, below left ] {\small{$ 1 $}};
    \draw [blue, fill = blue] ( 2, - 2) circle [radius = 0.15] node [black, below left ] {\small{$ 2 $}};
    \draw [blue, fill = blue] ( 3, - 3) circle [radius = 0.15] node [black, below left ] {\small{$ 3 $}};
    \draw [blue, fill = blue] ( 4, - 4) circle [radius = 0.15] node [black, below left ] {\small{$ 4 $}};
    \draw [blue, fill = blue] ( 5, - 5) circle [radius = 0.15] node [black, below left ] {\small{$ 5 $}};
    \draw [blue, fill = blue] ( 6, - 6) circle [radius = 0.15] node [black, below left ] {\small{$ 6 $}};
    \draw [blue, fill = blue] ( 7, - 7) circle [radius = 0.15] node [black, below left ] {\small{$ 7 $}};
    \draw [blue, fill = blue] ( 8, - 8) circle [radius = 0.15] node [black, below left ] {\small{$ 8 $}};
    \draw [blue, fill = blue] ( 9, - 9) circle [radius = 0.15] node [black, below left ] {\small{$ 9 $}};
    \draw [blue, fill = blue] (10, -10) circle [radius = 0.15] node [black, below left ] {\small{$ 10 $}};
    \draw [blue, fill = blue] (11, - 9) circle [radius = 0.15] node [black, below right] {\small{$ 11 $}};
    \draw [blue, fill = blue] (12, - 8) circle [radius = 0.15] node [black, below right] {\small{$ 12 $}};
    \draw [blue, fill = blue] (13, - 7) circle [radius = 0.15] node [black, below right] {\small{$ 13 $}};
    \draw [blue, fill = blue] (14, - 6) circle [radius = 0.15] node [black, below right] {\small{$ 14 $}};
    \draw [blue, fill = blue] (15, - 5) circle [radius = 0.15] node [black, below right] {\small{$ 15 $}};
    \draw [blue, fill = blue] (16, - 4) circle [radius = 0.15] node [black, below right] {\small{$ 16 $}};
    \draw [blue, fill = blue] (17, - 3) circle [radius = 0.15] node [black, below right] {\small{$ 17 $}};
    \draw [blue, fill = blue] (18, - 2) circle [radius = 0.15] node [black, below right] {\small{$ 18 $}};
    \draw [blue, fill = blue] (19, - 1) circle [radius = 0.15] node [black, below right] {\small{$ 19 $}};
    \draw [blue, fill = blue] (20,   0) circle [radius = 0.15] node [black, below right] {\small{$ 20 $}};
  \end{tikzpicture}
  \caption
  {
    Example of a polyline with two straight segments having extra vertices. The total number of vertices is $ 21 $.
  }
  \label{fig:ExampleTwoSegments}
\end{figure}
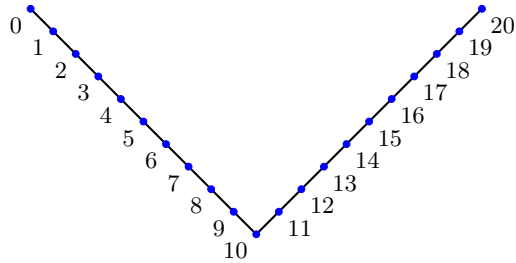

For a given source polyline, the task is to find the resultant polyline with the minimum number of segments and arcs within the specified tolerance and, among them, one with the minimum sum of squared deviations. This task is solved by using the dynamic programming approach (see \cite{OptimalCompressionWithArcs}, \cite{DynamicCompressionWithArcs}, and \cite[Sect. III]{OptimalCompressionOfAPolylineWithSegmentsAndArcs}). This method finds the minimum number of segments and arcs, while satisfying the tolerance requirement (or other criteria) for all parts of the source polyline between the first vertex and any other vertex; however, the resultant polyline will only include some of the vertices of the source polyline. For example, see Fig.~\ref{fig:ExampleTwoSegments} and assume that the distance between neighboring vertices is $ 3 $, then with a tolerance of~$ 2 $, the resultant polyline will have only two segments. From this example, it is obvious that the resultant polyline cannot have vertices~$ \overline{1..8} $ or $ \overline{12..19} $; otherwise, it will have more than two segments. However, optimal solutions for parts of the polyline from vertex~$ 0 $ to vertices~$ \overline{1..8} $ and $ \overline{12..19} $ are found. This leads to a higher complexity of the algorithm and prevents the practical application of such an algorithm to polylines with a~large number of vertices in arcs, because the task of fitting an arc to a set of $ n $ points within the specified tolerance has $ O{ \left( n \log{ \left( n \right) } \right) } $ complexity. The purpose of this paper is to develop another approach that avoids this inefficiency.

\section{Algorithm\label{section:algorithm}}

The algorithm is based on the dynamic programming approach; however, instead of processing at each iteration from the second to the last vertices of the source polyline, the algorithm finds the farthest vertex it can reach from the constructed solutions and analyzes backward if this solution is valid. Iterations from the second to the last vertices are replaced by the iterations from the first to the last segments and arcs. The algorithm terminates when the end vertex of the source polyline is reached.

Define a polyline as an ordered set of vertices $ p_i $, $ i = \overline{0..n - 1} $, where n is the number of vertices in the source polyline. Define the array $ \DeclareForwardSegment $ for the indices, satisfying that no segments can be fitted for the part of the source polyline between the vertex $ i $ and any vertices after $ \DeclareForwardSegment_i $ within the specified tolerance. Similarly, define the array $ \DeclareForwardArc $ for the fitting of arcs. The algorithm for arcs is described in \cite[Appendix~III]{OptimalCompressionOfAPolylineWithSegmentsAndArcs}. A similar algorithm based on the convex hull can be used for segments \cite{OptimalCompressionOfAPolylineWithSegmentsAndArcs}. Define the array $ \DeclareBackwardSegment $ for the indices, satisfying that no segments can be fitted between any vertex before $ \DeclareBackwardSegment_i $ and the vertex $ i $ within the specified tolerance. Similarly, define the array $ \DeclareBackwardArc $ for the fitting of arcs. Note that these arrays are not required to contain the minimum possible values for $ \DeclareForwardSegment $ and $ \DeclareForwardArc $ or the maximum possible values for $ \DeclareBackwardSegment $ and $ \DeclareBackwardArc $, because the algorithm described in this section will produce the same result as long as the definitions for these arrays are satisfied. Nevertheless, the higher quality of these arrays will lead to a faster algorithm.

Let the array $ \mathrm{LAST} $ store the last possible position for the penalty $ p $, so that any polyline from vertex $ 0 $ to any vertex after $ \mathrm{LAST}_p $ cannot have penalty $ p $. The array is initialized from position $ 0 $ to $ \mathrm{processed} $.

Following \cite{OptimalCompressionOfAPolylineWithSegmentsAndArcs}, the penalty for each segment will be $ \DeclarePenaltySegment = 2 $, and the penalty for each arc will be $ \DeclarePenaltyArc = 3 $.\footnote{Other penalties can be used instead; however, if they are natural numbers, no changes in the algorithm are necessary. Obviously, any penalties for segments and arcs as rational numbers can be converted to equivalent penalties in natural numbers.} Let the array $ \mathrm{PENALTY} $ store the possible or found penalty for the $ i $-th vertex; the boolean array $ \mathrm{SOLVED} $ store $ \mathrm{FALSE} $ for possible solutions and $ \mathrm{TRUE} $ for found solutions; and if the solution is found, the array $ \mathrm{ERROR} $\footnote{Instead of using the array $ \mathrm{SOLVED} $ to distinguish possible from found solutions, the array $ \mathrm{ERROR} $ can store $ \infty $ for possible solutions and the minimum sum of the squared deviations for the found solutions.} will store the minimum sum of the squared deviations from vertices of the source polyline to the resultant polyline\footnote{The sum of the squared deviations can be replaced by an integral. In the case of arcs, the sum or integral of squared deviations can be approximate (see \cite{FittingOfCircularArcsWithO1Complexity} and \cite{EfficientFittingOfCircularArcs}).}. These arrays are initialized from position $ 0 $ to $ \mathrm{position} $.

To simplify the pseudocode, assume that $ \forall i < 0 $ $ \Rightarrow $ \\ $ \DeclareForwardSegment_i = -\infty $, $ \DeclareForwardArc_i = -\infty $, and $ \mathrm{LAST}_i = -\infty $. Algorithm~\ref{fig:FindOptimalSolution} shows the pseudocode for finding the optimal solution. This is the maximum jump step. Instead of solving from the second to the last vertices, the algorithm tries to find the farthest vertex with the next penalty ($ \mathrm{processed} + 1 $)\footnote{If other penalties are assigned for segments and arcs, then it might be more efficient to find the next penalty by analyzing the last entries of the array $ \mathrm{LAST} $.} until the solution for the whole source polyline is found. The algorithm jumps from the known solution with the help of $ \DeclareForwardArc $ for arcs or $ \DeclareForwardSegment $ for segments. The jump can be longer than it should be, but it can never be shorter, which guarantees that the optimal solution is not missed. The preference is given to arcs because they have a higher penalty, which leads to using a smaller penalty to jump from. To guarantee that the jump is performed from a known solution, the function $ Solve $ is called. Note that when the algorithm progresses farther by evaluating possible solutions, the array $ \mathrm{LAST} $ does not necessarily point to the proper position; therefore, the function $ \CheckAndAdjustLastPosition $ is used to make decremental adjustments. When the jump reaches the last vertex of the source polyline, the last vertex is checked if it can be solved for that penalty ($ processed + 1 $). If it can, then the solution is found.

\begin{algorithm} [p]
  \caption
  {
    Algorithm to find the optimal compression of a polyline with segments and arcs.
  }
  \label{fig:FindOptimalSolution}
  \begin{algorithmic} 
    \Function{FindOptimalSolution}{}
      \State $ \mathrm{LAST}_0 = 0 $, $ \mathrm{processed} = 0 $
      \State $ \mathrm{PENALTY}_0 = 0 $, $ \mathrm{SOLVED}_0 = \mathrm{TRUE} $, $ \mathrm{ERROR}_0 = 0 $, $ \mathrm{position} = 0 $
      \Label{TryNextPenalty:}
        \State $ \mathrm{processed} = \mathrm{processed} + 1 $, $ \mathrm{LAST}_{\mathrm{processed}} = -\infty $
        \While{ $ \mathrm{TRUE} $ }
          \State $ i = \DeclareForwardSegment_{\mathrm{LAST}_{\mathrm{processed} - \DeclarePenaltySegment}} $,
          $ j = \DeclareForwardArc_{\mathrm{LAST}_{\mathrm{processed} - \DeclarePenaltyArc}} $
          \State $ k = \max{ \left( i, j \right) } $
          \If{ $ k = -\infty $ }
            \Goto{ TryNextPenalty }
          \EndIf
          \If{ $ i \leq j $ }
            \If{ \Not \Call{\CheckAndAdjustLastPosition}{$ \mathrm{processed} - \DeclarePenaltyArc $} }
              \Continue
            \EndIf
          \EndIf
          \If{ $ i \geq j $ }
            \If { \Not \Call{\CheckAndAdjustLastPosition}{$ \mathrm{processed} - \DeclarePenaltySegment $} }
              \Continue
            \EndIf
          \EndIf
          \If{ $ i < j $ }
            \If{ \Not \Call{Solve}{$ j $, $ \mathrm{processed} - \DeclarePenaltyArc $} }
              \Continue
            \EndIf
          \ElsIf{$ i > j $}
            \If{ \Not \Call{Solve}{$ i $, $ \mathrm{processed} - \DeclarePenaltySegment $} }
              \Continue
            \EndIf
          \Else
            \If{ \Not $ \mathrm{SOLVED}_{\mathrm{LAST}_{\mathrm{processed} - \DeclarePenaltySegment}} $ \MyAnd
            \State \Not $ \mathrm{SOLVED}_{\mathrm{LAST}_{\mathrm{processed} - \DeclarePenaltyArc}} $ }
              \If{ \Not \Call{Solve}{$ j $, $ \mathrm{processed} - \DeclarePenaltyArc $} }
                \Continue
              \EndIf
            \EndIf
          \EndIf
        \EndWhile
        \While{$ \mathrm{position} < k $}
          \State $ \mathrm{position} = \mathrm{position} + 1 $
          \State $ \mathrm{PENALTY}_{\mathrm{position}} = \mathrm{processed} $, $ \mathrm{SOLVED}_{\mathrm{position}} = \mathrm{FALSE} $
        \EndWhile
        \State $ \mathrm{LAST}_{\mathrm{processed}} = \mathrm{position} $
        \If{ $ \mathrm{position} + 1 = n $ }
          \If{ \Call{Solve}{$ \mathrm{position} $, $ \mathrm{processed} $} }
            \Return
          \EndIf
          \State $ \mathrm{LAST}_{\mathrm{processed}} = \mathrm{LAST}_{\mathrm{processed}} - 1 $
        \EndIf
      \State \Goto{ TryNextPenalty }
    \EndFunction
    \State
    \Function{\CheckAndAdjustLastPosition}{$ \mathrm{penalty} $}
      \If{ $ \mathrm{SOLVED}_{\mathrm{LAST}_{\mathrm{penalty}}} $ \MyAnd $ \mathrm{PENALTY}_{\mathrm{LAST}_{\mathrm{penalty}}} = \mathrm{penalty} $ \MyOr 
          \State \Not $ \mathrm{SOLVED}_{\mathrm{LAST}_{\mathrm{penalty}}} $ \MyAnd $ \mathrm{PENALTY}_{\mathrm{LAST}_{\mathrm{penalty}}} \leq \mathrm{penalty} $ }
        \State \Return $ \mathrm{TRUE} $
      \EndIf
      \State $ \mathrm{LAST}_{\mathrm{penalty}} = \mathrm{LAST}_{\mathrm{penalty}} - 1 $
      \State \Return $ \mathrm{FALSE} $
    \EndFunction
  \end{algorithmic}
\end{algorithm}

Algorithm~\ref{fig:Solve} recursively finds the solution for the specified position and penalty. This is the looking backward step. It returns $ \mathrm{TRUE} $ if such a solution is found and $ \mathrm{FALSE} $ otherwise. When the algorithm searches backward for the range of possible beginnings of the segment or arc, the array $ \mathrm{LAST} $ does not necessarily point to the proper position; therefore, the function $ Adjust $ is used to iteratively decrement its values. Note that the array $ \mathrm{PENALTY} $ might also contain values that contradict the values in the array $ \mathrm{LAST} $. This happens when the solution does not exist for some values in the array $ \mathrm{PENALTY} $ and, as a result, some of its values are incremented.

\begin{algorithm} [p]
  \caption
  {
    Recursive algorithm to find the solution at the position $ i $ for the specified $ penalty $.
  }
  \label{fig:Solve}
  \begin{algorithmic} 
    \Function{Solve}{$ i, \mathrm{penalty} $}
      \If{ $ \mathrm{SOLVED}_{i} $ }
        \Return $ \mathrm{PENALTY}_{i} = \mathrm{penalty} $
      \EndIf
      \While{ $ \mathrm{PENALTY}_{i} \leq \mathrm{penalty} $ }
        \If{ $ i \leq \mathrm{LAST}_{\mathrm{PENALTY}_{i}} $}
          \If{ $ \DeclarePenaltyArc \leq \mathrm{PENALTY}_{i} $ }
            \State $ a_0 = \DeclareBackwardArc_{i} $
            \State \Call{Adjust}{ $ a_0 $, $ \mathrm{PENALTY}_{i} - \DeclarePenaltyArc $ }
            \State $ a_1 = \min{ \left( \mathrm{LAST}_{\mathrm{PENALTY}_{i} - \DeclarePenaltyArc}, i - 3 \right) } $
            \State \Comment{If $ 4 $ is the minimum number of vertices in an arc.}
            \For{ $ k $ \MyFrom $ a_1 $ \MyTo $ a_0 $ \MyStep $ -1 $ }
              \If{ \Call{Solve}{ $ k $, $ \mathrm{PENALTY}_{i} - \DeclarePenaltyArc $ } }
                \If{ an arc can be fitted to the part of the polyline between
                \State vertex $ k $ and vertex $ i $ }
                  \State $ \mathrm{new\_error} = \mathrm{ERROR}_{k} + $ the sum of squared
                  \State deviations for the fitted arc (can be approximate).
                  \If{ \Not $ \mathrm{SOLVED}_{i} $ }
                    \State $ \mathrm{ERROR}_i = \mathrm{new\_error} $
                    \State $ \mathrm{SOLVED}_{i} = \mathrm{TRUE} $
                  \Else
                    \State $ \mathrm{ERROR}_i = \min{ \left( \mathrm{ERROR}_i, \mathrm{new\_error} \right) } $
                  \EndIf
                \EndIf
              \EndIf
            \EndFor
          \EndIf
          \If{ $ \DeclarePenaltySegment \leq \mathrm{PENALTY}_{i} $ }
            \State $ s_0 = \DeclareBackwardSegment_{i} $
            \State \Call{Adjust}{ $ s_0 $, $ \mathrm{PENALTY}_{i} - \DeclarePenaltySegment $ }
            \State $ s_1 = \min{ \left( \mathrm{LAST}_{\mathrm{PENALTY}_{i} - \DeclarePenaltySegment}, i - 1 \right) } $
            \For{ $ k $ \MyFrom $ s_1 $ \MyTo $ s_0 $ \MyStep $ -1 $ }
              \If{ \Call{Solve}{ $ k $, $ \mathrm{PENALTY}_{i} - \DeclarePenaltySegment $ } }
                \If{ a segment can be fitted to the part of the polyline
                \State between vertex $ k $ and vertex $ i $ }
                  \State $ \mathrm{new\_error} = \mathrm{ERROR}_{k} + $ the sum of squared
                  \State deviations for the fitted segment.
                  \If{ \Not $ \mathrm{SOLVED}_{i} $ }
                    \State $ \mathrm{ERROR}_i = \mathrm{new\_error} $
                    \State $ \mathrm{SOLVED}_{i} = \mathrm{TRUE} $
                  \Else
                    \State $ \mathrm{ERROR}_i = \min{ \left( \mathrm{ERROR}_i, \mathrm{new\_error} \right) } $
                  \EndIf
                \EndIf
              \EndIf
            \EndFor
          \EndIf
        \EndIf
        \If{ $ \mathrm{SOLVED}_{i} $ }
          \Return $ \mathrm{PENALTY}_{i} = \mathrm{penalty} $
        \EndIf
        \State $ \mathrm{PENALTY}_{i} = \mathrm{PENALTY}_{i} + 1 $
      \EndWhile
      \State \Return $ \mathrm{FALSE} $
    \EndFunction
    \State
    \Function{Adjust}{$ i, \mathrm{penalty} $}
      \While{ $ i \leq \mathrm{LAST}_{\mathrm{penalty}} $ }
        \If{ \Call{\CheckAndAdjustLastPosition}{$ \mathrm{penalty} $} }
          \Return
        \EndIf
      \EndWhile
    \EndFunction
  \end{algorithmic}
\end{algorithm}

In this approach, segments or arcs are fitted to any part of the source polyline no more than once. This is guaranteed because function $ Solve $ only fits segments or arcs when the solution for the starting vertex is found and penalty plus $ \DeclarePenaltySegment $ for segments or $ \DeclarePenaltyArc $ for arcs equals $ \mathrm{PENALTY}_i $ in Algorithm~\ref{fig:Solve}.

If the noise in the vertices of the source polyline does not introduce backward movement along corresponding true arcs, it is possible to use the approach from \cite{OptimalComressionWithMinimumComplexity}. This will further reduce the complexity of the algorithm for the part where multiple arcs have to be fitted from the same point $ i $ in Algorithm~\ref{fig:Solve}.

\section{Performance Evaluation\label{section:performance}}

The performance of the algorithm was evaluated on $ 100 $~arcs of radius $ 1 $, following each other, and added uniform radial noise of $ 0.05 $, shown in Fig.~\ref{fig:ExampleArcs}. The tolerance was set to $ 0.06 $. On this data, the algorithm performs similarly to or better than the algorithm based on the dynamic programming approach (see the comparison in Fig.~\ref{fig:PerformanceEvaluationArcs}). The calculation was performed using the Intel Xeon CPU $\text{E5-2670}$ processor. The advantage becomes clear when there is a large number of points in each arc. Note that the nonsmooth behavior in both graphs can be explained by the same recursive algorithm used in the preparation step to check where arcs can be fitted in both algorithms (see \cite[Appendix~III]{OptimalCompressionOfAPolylineWithSegmentsAndArcs}).

\begin{figure} [!t]

\centering
\begin{tikzpicture} [scale = 0.5*2]
  \draw [thick, black]
    (1, 1) -- 
    (1.03777, 1.04727) -- 
    (1.041, 1.09445) -- 
    (1.01565, 1.14601) -- 
    (1.06619, 1.18575) -- 
    (1.04443, 1.23936) -- 
    (1.0037, 1.30223) -- 
    (1.06121, 1.3359) -- 
    (1.11544, 1.3664) -- 
    (1.0897, 1.43054) -- 
    (1.10615, 1.47777) -- 
    (1.17749, 1.493) -- 
    (1.16386, 1.55869) -- 
    (1.17353, 1.61295) -- 
    (1.24851, 1.61673) -- 
    (1.26508, 1.6661) -- 
    (1.31059, 1.68941) -- 
    (1.34242, 1.72553) -- 
    (1.34637, 1.79645) -- 
    (1.40581, 1.80117) -- 
    (1.45721, 1.81234) -- 
    (1.4969, 1.83938) -- 
    (1.51687, 1.90388) -- 
    (1.57424, 1.9002) -- 
    (1.62473, 1.90597) -- 
    (1.66911, 1.92476) -- 
    (1.72094, 1.91992) -- 
    (1.76627, 1.93311) -- 
    (1.81332, 1.93853) -- 
    (1.85041, 2.00844) -- 
    (1.90053, 2.00989) -- 
    (1.94951, 2.02778) -- 
    (2, 1.98925) -- 
    (2.04921, 2.00179) -- 
    (2.09702, 1.98507) -- 
    (2.14219, 1.95855) -- 
    (2.19698, 1.9903) -- 
    (2.25242, 2.00771) -- 
    (2.28653, 1.94456) -- 
    (2.32134, 1.89809) -- 
    (2.37439, 1.90386) -- 
    (2.44403, 1.93882) -- 
    (2.46017, 1.86091) -- 
    (2.51516, 1.85949) -- 
    (2.52888, 1.79152) -- 
    (2.5961, 1.80374) -- 
    (2.66604, 1.81158) -- 
    (2.67886, 1.74901) -- 
    (2.73754, 1.73754) -- 
    (2.76854, 1.69656) -- 
    (2.74689, 1.61296) -- 
    (2.82696, 1.61331) -- 
    (2.831, 1.55525) -- 
    (2.86208, 1.51671) -- 
    (2.9021, 1.48218) -- 
    (2.86006, 1.40678) -- 
    (2.91924, 1.38076) -- 
    (2.91172, 1.32622) -- 
    (2.91452, 1.27742) -- 
    (2.94014, 1.23549) -- 
    (2.93295, 1.18557) -- 
    (2.94808, 1.14063) -- 
    (2.94549, 1.09312) -- 
    (2.96226, 1.04727) -- 
    (3, 1) -- 
    (3.02807, 1.04775) -- 
    (2.96389, 1.10205) -- 
    (3.00326, 1.14785) -- 
    (2.97764, 1.20336) -- 
    (3.0761, 1.23142) -- 
    (3.01771, 1.29797) -- 
    (3.03499, 1.34529) -- 
    (3.05177, 1.39277) -- 
    (3.052, 1.44837) -- 
    (3.11239, 1.47444) -- 
    (3.14351, 1.51336) -- 
    (3.18965, 1.54146) -- 
    (3.15973, 1.62319) -- 
    (3.21455, 1.6446) -- 
    (3.256, 1.67433) -- 
    (3.28603, 1.71397) -- 
    (3.3236, 1.74629) -- 
    (3.35367, 1.78756) -- 
    (3.39115, 1.82094) -- 
    (3.44668, 1.82811) -- 
    (3.4789, 1.8694) -- 
    (3.51566, 1.90615) -- 
    (3.55548, 1.93987) -- 
    (3.61069, 1.93987) -- 
    (3.66779, 1.92848) -- 
    (3.7239, 1.91018) -- 
    (3.76569, 1.9354) -- 
    (3.80502, 1.98024) -- 
    (3.84991, 2.0118) -- 
    (3.89923, 2.02311) -- 
    (3.95206, 1.97591) -- 
    (4, 1.96407) -- 
    (4.05041, 2.02608) -- 
    (4.09776, 1.99258) -- 
    (4.15358, 2.03534) -- 
    (4.20274, 2.01925) -- 
    (4.2474, 1.98766) -- 
    (4.29432, 1.97026) -- 
    (4.34968, 1.97729) -- 
    (4.39283, 1.94837) -- 
    (4.40802, 1.86269) -- 
    (4.48868, 1.91426) -- 
    (4.50219, 1.83786) -- 
    (4.5808, 1.86923) -- 
    (4.5701, 1.76868) -- 
    (4.62092, 1.7566) -- 
    (4.66553, 1.7343) -- 
    (4.72252, 1.72252) -- 
    (4.75194, 1.68152) -- 
    (4.78004, 1.64016) -- 
    (4.77062, 1.57153) -- 
    (4.7983, 1.53341) -- 
    (4.83476, 1.50033) -- 
    (4.89963, 1.48086) -- 
    (4.89075, 1.42129) -- 
    (4.92359, 1.38257) -- 
    (4.95709, 1.34245) -- 
    (4.97267, 1.29506) -- 
    (4.9514, 1.23831) -- 
    (4.95376, 1.18972) -- 
    (5.01924, 1.15119) -- 
    (5.01908, 1.10037) -- 
    (4.96402, 1.04736) -- 
    (5, 1) -- 
    (4.97257, 1.05047) -- 
    (5.00491, 1.09801) -- 
    (5.03414, 1.14327) -- 
    (5.0188, 1.19517) -- 
    (5.03063, 1.24282) -- 
    (5.05535, 1.28656) -- 
    (5.05715, 1.33736) -- 
    (5.10135, 1.37223) -- 
    (5.11565, 1.41826) -- 
    (5.126, 1.46716) -- 
    (5.17445, 1.49482) -- 
    (5.14274, 1.5728) -- 
    (5.16819, 1.61691) -- 
    (5.19831, 1.65793) -- 
    (5.29198, 1.64171) -- 
    (5.2949, 1.7051) -- 
    (5.31425, 1.75661) -- 
    (5.39589, 1.73611) -- 
    (5.38149, 1.83396) -- 
    (5.43564, 1.84463) -- 
    (5.47443, 1.87685) -- 
    (5.53475, 1.87043) -- 
    (5.59178, 1.8631) -- 
    (5.61681, 1.9251) -- 
    (5.65926, 1.95229) -- 
    (5.7136, 1.94412) -- 
    (5.75958, 1.95982) -- 
    (5.80112, 1.99985) -- 
    (5.85797, 1.95749) -- 
    (5.90233, 1.99168) -- 
    (5.95047, 2.00813) -- 
    (6, 1.99594) -- 
    (6.04673, 1.9512) -- 
    (6.09784, 1.99342) -- 
    (6.14573, 1.98246) -- 
    (6.19113, 1.96087) -- 
    (6.23311, 1.93061) -- 
    (6.28529, 1.94047) -- 
    (6.32445, 1.90676) -- 
    (6.37743, 1.91119) -- 
    (6.44153, 1.93353) -- 
    (6.47276, 1.88447) -- 
    (6.49005, 1.81761) -- 
    (6.58009, 1.86817) -- 
    (6.57844, 1.77994) -- 
    (6.62316, 1.75932) -- 
    (6.67348, 1.74307) -- 
    (6.73723, 1.73723) -- 
    (6.74677, 1.67683) -- 
    (6.77424, 1.6354) -- 
    (6.79314, 1.58823) -- 
    (6.81252, 1.54291) -- 
    (6.85288, 1.5112) -- 
    (6.89355, 1.47762) -- 
    (6.90264, 1.42692) -- 
    (6.96158, 1.3983) -- 
    (6.95459, 1.34156) -- 
    (6.96466, 1.29263) -- 
    (6.96906, 1.24274) -- 
    (6.95225, 1.18941) -- 
    (6.96053, 1.14248) -- 
    (6.99385, 1.09789) -- 
    (7.03618, 1.0509) -- 
    (7, 1) -- 
    (6.95995, 1.05109) -- 
    (7.02749, 1.09578) -- 
    (7.01553, 1.14603) -- 
    (7.01287, 1.19635) -- 
    (7.05779, 1.23601) -- 
    (7.08631, 1.27717) -- 
    (7.03862, 1.34399) -- 
    (7.06398, 1.38771) -- 
    (7.09768, 1.42677) -- 
    (7.10801, 1.47678) -- 
    (7.18009, 1.49144) -- 
    (7.17602, 1.55057) -- 
    (7.17654, 1.61072) -- 
    (7.25751, 1.60935) -- 
    (7.24384, 1.68534) -- 
    (7.32226, 1.67774) -- 
    (7.35775, 1.70862) -- 
    (7.38556, 1.7487) -- 
    (7.38427, 1.83022) -- 
    (7.4394, 1.839) -- 
    (7.46578, 1.89128) -- 
    (7.50733, 1.92173) -- 
    (7.59295, 1.86063) -- 
    (7.61322, 1.93376) -- 
    (7.66309, 1.94159) -- 
    (7.71194, 1.9496) -- 
    (7.7512, 1.99327) -- 
    (7.80994, 1.9555) -- 
    (7.85305, 1.99065) -- 
    (7.90192, 1.99587) -- 
    (7.9511, 1.99535) -- 
    (8, 1.99495) -- 
    (8.04732, 1.96313) -- 
    (8.09884, 2.00354) -- 
    (8.1491, 2.00513) -- 
    (8.19369, 1.97374) -- 
    (8.24242, 1.9678) -- 
    (8.29121, 1.95999) -- 
    (8.32883, 1.91903) -- 
    (8.3982, 1.96133) -- 
    (8.41729, 1.88229) -- 
    (8.4801, 1.8982) -- 
    (8.52097, 1.86919) -- 
    (8.57865, 1.86601) -- 
    (8.62281, 1.83976) -- 
    (8.65249, 1.79507) -- 
    (8.68517, 1.75597) -- 
    (8.69859, 1.69859) -- 
    (8.73939, 1.67014) -- 
    (8.76385, 1.62688) -- 
    (8.81368, 1.60346) -- 
    (8.84444, 1.56423) -- 
    (8.89974, 1.53928) -- 
    (8.8707, 1.4654) -- 
    (8.92811, 1.43897) -- 
    (8.89121, 1.36915) -- 
    (8.94825, 1.33929) -- 
    (8.93141, 1.28254) -- 
    (8.94935, 1.2378) -- 
    (9.00032, 1.19898) -- 
    (9.01932, 1.1512) -- 
    (9.01762, 1.10023) -- 
    (8.9891, 1.04859) -- 
    (9, 1);
\end{tikzpicture}
  \caption
  {
    Example of four arcs of radius $ 1 $ with $ 64 $ points in each.
  }
  \label{fig:ExampleArcs}

\vspace{0.5 cm}

  \centering
  \pgfplotsset{width = 0.99\columnwidth}
  \begin{tikzpicture}
    \begin{axis}
      [
        xlabel = {Number of points in each arc},
        ylabel = {Time (seconds)},
        xmin = 0, 
        xmax = 256,
        ymin = 0, 
        ymax = 6,
        xtick = {8, 16, 32, 64, 128, 256},
        xticklabels={8, , 32, 64, 128, 256, 512, 1024},
        ytick = {0, 1, 2, 3, 4, 5, 6},
        legend pos = north west,
        xmajorgrids = true,
        ymajorgrids = true,
        grid style = dashed,
      ]

      \addplot+
        [
          color = blue,
          mark options = {mark = square*, fill = blue}
        ]
        coordinates 
        {
          (8, 0.08121133) (10, 0.13252382) (11, 0.12898202) (13, 0.14803834) (16, 0.13074975) (19, 0.21248546) (23, 0.23791861) (27, 0.34516479) (32, 0.31922513) (38, 0.42040159) (45, 0.51034139) (54, 0.67690935) (64, 0.66530593) (76, 0.98364086) (91, 1.25367140) (108, 1.58111675) (128, 1.62600926) (152, 2.77985967) (181, 3.91181570) (215, 5.03750946) (256, 5.98500895)
        };

      \addplot+
        [
          color = green,
          mark options = {mark = square*, fill = green}
        ]
        coordinates 
        {
          (8, 0.08180787) (10, 0.13322353) (11, 0.12751298) (13, 0.14401736) (16, 0.12656996) (19, 0.20647489) (23, 0.23151170) (27, 0.33513815) (32, 0.30350334) (38, 0.39500967) (45, 0.47929584) (54, 0.64454531) (64, 0.62483326) (76, 0.88755295) (91, 1.10804684) (108, 1.41851196) (128, 1.39440442) (152, 2.28594751) (181, 3.20149372) (215, 4.10237349) (256, 4.55180197)
        };
    \end{axis}
  \end{tikzpicture}
  \caption
  {
    Average time to fit $ 100 $~arcs (see Fig.~\ref{fig:ExampleArcs}) using the dynamic programming approach (blue) and the algorithm described in this paper (green).
  }
  \label{fig:PerformanceEvaluationArcs}
\end{figure}

\begin{figure} [!t]
  \input{ExamplePolylineSegments.tex}
  \caption
  {
    Example of eight segments of length $ 1 $ with $ 64 $ points in each.
  }
  \label{fig:ExampleSegments}

\vspace{0.5 cm}

  \centering
  \pgfplotsset{width = 0.99\columnwidth}
  \begin{tikzpicture}
    \begin{axis}
      [
        xlabel = {Number of points in each segment},
        ylabel = {Time (seconds)},
        xmin = 0, 
        xmax = 256,
        ymin = 0, 
        ymax = 1.1,
        xtick = {8, 16, 32, 64, 128, 256},
        xticklabels={8, , 32, 64, 128, 256, 512, 1024},
        ytick = {0, 0.2, 0.4, 0.6, 0.8, 1.0},
        legend pos = north west,
        xmajorgrids = true,
        ymajorgrids = true,
        grid style = dashed,
      ]
      
      \addplot+
        [
          color = blue,
          mark options = {mark = square*, fill = blue}
        ]
        coordinates 
        {
          (8, 0.01006073) (10, 0.01370227) (11, 0.01515984) (13, 0.01838282) (16, 0.02492679) (19, 0.03122963) (23, 0.04078470) (27, 0.05091773) (32, 0.06049855) (38, 0.07513088) (45, 0.09447850) (54, 0.12101356) (64, 0.15530176) (76, 0.18699581) (91, 0.23691843) (108, 0.29698661) (128, 0.39598250) (152, 0.48408530) (181, 0.61892235) (215, 0.78715461) (256, 1.05772234)
        };

      \addplot+
        [
          color = green,
          mark options = {mark = square*, fill = green}
        ]
        coordinates 
        {
          (8, 0.01048816) (10, 0.01358262) (11, 0.01499308) (13, 0.01828418) (16, 0.02588902) (19, 0.02948170) (23, 0.03702688) (27, 0.04468847) (32, 0.05634687) (38, 0.06636559) (45, 0.08085345) (54, 0.09730157) (64, 0.11841675) (76, 0.14091285) (91, 0.17305546) (108, 0.20826207) (128, 0.24896255) (152, 0.29846708) (181, 0.35892897) (215, 0.43545225) (256, 0.52377327)
        };
    \end{axis}
  \end{tikzpicture}
  \caption
  {
    Average time to fit $ 100 $~segments (see Fig.~\ref{fig:ExampleSegments}) using the dynamic programming approach (blue) and the algorithm described in this paper (green).
  }
  \label{fig:PerformanceEvaluationArcsSegments}
\end{figure}

This algorithm is also applicable for optimal compression of a polyline with segments. The performance was evaluated on the polyline with $ 100 $~segments of length $ 1 $, forming a zigzag line, and added uniform perpendicular noise of $ 0.05 $, shown in Fig.~\ref{fig:ExampleSegments}. The tolerance was set to $ 0.06 $. Using this data, this algorithm also performs better than the algorithm based on the dynamic programming approach (see Fig.~\ref{fig:PerformanceEvaluationArcsSegments}).

\section{Example}

\begin{figure*} [htb]
\centering
\includegraphics[width = \textwidth, keepaspectratio]{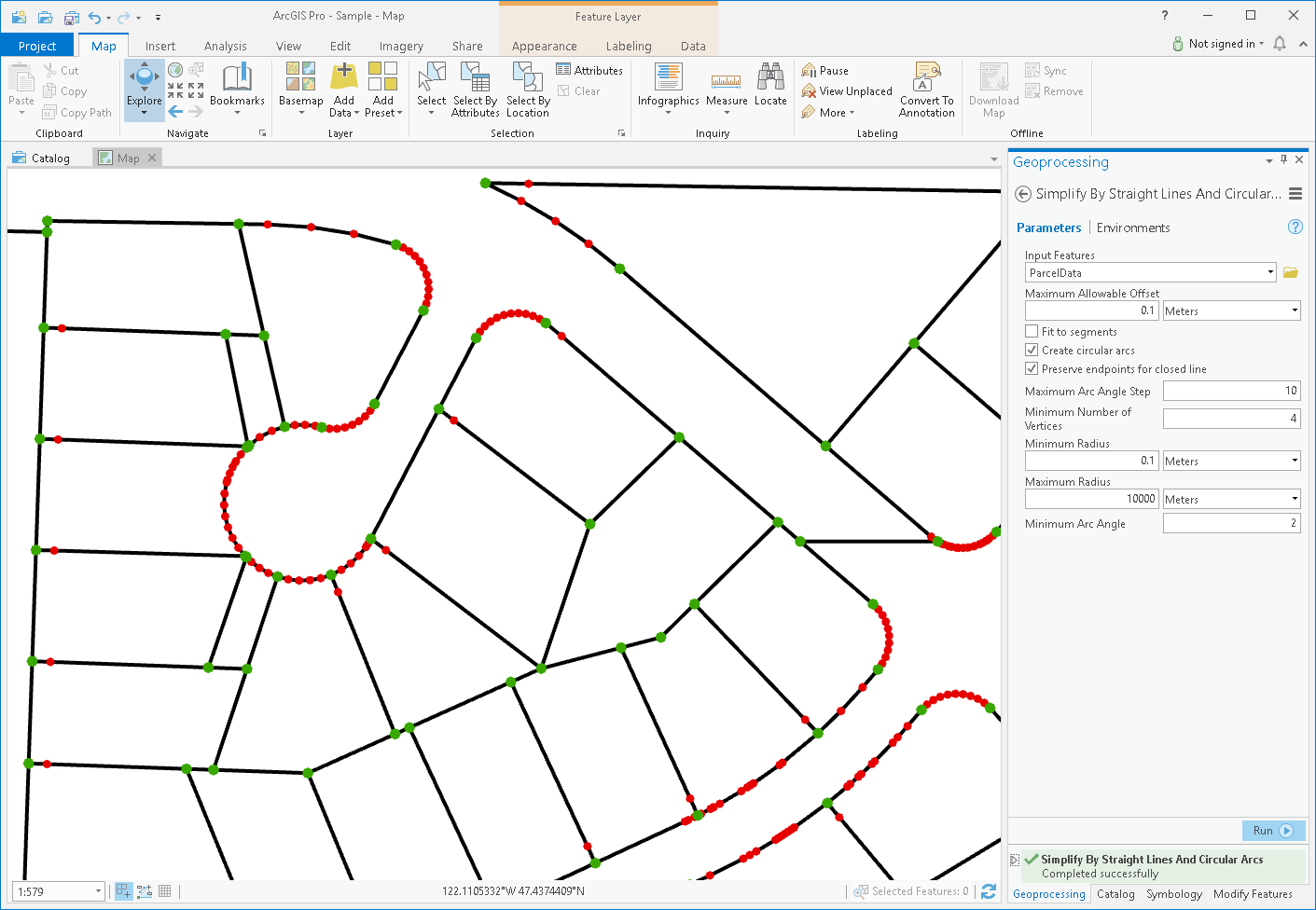}
\caption
{
  Example of parcel data with lost circular arcs. The ArcGIS Pro geoprocessing tool Simplify by Straight Lines and Circular Arcs was applied to this data. The black lines are the source polylines, the red circles are vertices of the source polylines, and the green circles are resultant vertices.
}
\label{fig:ExampleArcGISProGPTool}
\end{figure*}

The algorithm described in this paper was implemented in the ArcGIS Pro geoprocessing tool Simplify by Straight Lines and Circular Arcs. This algorithm was applied to parcel data where original arcs were lost due to digitization, limitations of the format, projection, or some other reason (see Fig.~\ref{fig:ExampleArcGISProGPTool}). The restoration of original arcs is an important task because it creates cleaner databases and simplifies future editing.

\section{Conclusion}

The modification to the dynamic programming approach described in this paper produces a faster algorithm, especially in cases where arcs or segments have many points. However, if the source polyline does not have a clear structure of segments and arcs, for example, a polyline generated by a random walk, this algorithm will not be any faster than the dynamic programming approach. In the worst case, the algorithm would not be significantly slower than the dynamic programming approach; therefore, the algorithm is applicable to any input data.

\section{Future Work}

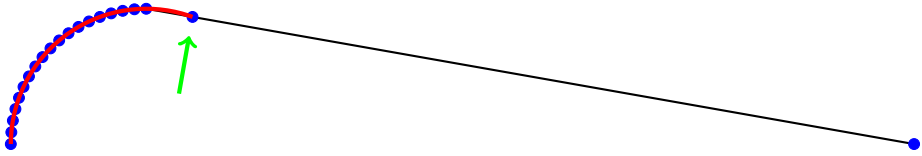
\begin{figure*} [!htb]
  \centering
  \begin{tikzpicture} [scale = 1.8]
    \draw [thick, black]
      (-1.00000000000000, 0.00000000000000) --
      (-0.99619469809175, 0.08715574274766) --
      (-0.98480775301221, 0.17364817766693) --
      (-0.96592582628907, 0.25881904510252) --
      (-0.93969262078591, 0.34202014332567) --
      (-0.90630778703665, 0.42261826174070) --
      (-0.86602540378444, 0.50000000000000) --
      (-0.81915204428899, 0.57357643635105) --
      (-0.76604444311898, 0.64278760968654) --
      (-0.70710678118655, 0.70710678118655) --
      (-0.64278760968654, 0.76604444311898) --
      (-0.57357643635105, 0.81915204428899) --
      (-0.50000000000000, 0.86602540378444) --
      (-0.42261826174070, 0.90630778703665) --
      (-0.34202014332567, 0.93969262078591) --
      (-0.25881904510252, 0.96592582628907) --
      (-0.17364817766693, 0.98480775301221) --
      (-0.08715574274766, 0.99619469809175) --
      (0.00000000000000, 1.00000000000000) --
      (0.34202014332567, 0.93969262078591) --
      (5.67128181961770, 0.00000000000000);

    \draw [ultra thick, green, ->] (0.242020143, 0.372564439) -- (0.317020143, 0.797910575);

    \draw [blue, fill = blue] (-1.00000000000000, 0.00000000000000) circle [radius = 0.04];
    \draw [blue, fill = blue] (-0.99619469809175, 0.08715574274766) circle [radius = 0.04];
    \draw [blue, fill = blue] (-0.98480775301221, 0.17364817766693) circle [radius = 0.04];
    \draw [blue, fill = blue] (-0.96592582628907, 0.25881904510252) circle [radius = 0.04];
    \draw [blue, fill = blue] (-0.93969262078591, 0.34202014332567) circle [radius = 0.04];
    \draw [blue, fill = blue] (-0.90630778703665, 0.42261826174070) circle [radius = 0.04];
    \draw [blue, fill = blue] (-0.86602540378444, 0.50000000000000) circle [radius = 0.04];
    \draw [blue, fill = blue] (-0.81915204428899, 0.57357643635105) circle [radius = 0.04];
    \draw [blue, fill = blue] (-0.76604444311898, 0.64278760968654) circle [radius = 0.04];
    \draw [blue, fill = blue] (-0.70710678118655, 0.70710678118655) circle [radius = 0.04];
    \draw [blue, fill = blue] (-0.64278760968654, 0.76604444311898) circle [radius = 0.04];
    \draw [blue, fill = blue] (-0.57357643635105, 0.81915204428899) circle [radius = 0.04];
    \draw [blue, fill = blue] (-0.50000000000000, 0.86602540378444) circle [radius = 0.04];
    \draw [blue, fill = blue] (-0.42261826174070, 0.90630778703665) circle [radius = 0.04];
    \draw [blue, fill = blue] (-0.34202014332567, 0.93969262078591) circle [radius = 0.04];
    \draw [blue, fill = blue] (-0.25881904510252, 0.96592582628907) circle [radius = 0.04];
    \draw [blue, fill = blue] (-0.17364817766693, 0.98480775301221) circle [radius = 0.04];
    \draw [blue, fill = blue] (-0.08715574274766, 0.99619469809175) circle [radius = 0.04];
    \draw [blue, fill = blue] (0.00000000000000, 1.00000000000000) circle [radius = 0.04];
    \draw [blue, fill = blue] (0.34202014332567, 0.93969262078591) circle [radius = 0.04];
    \draw [blue, fill = blue] (5.67128181961770, 0.00000000000000) circle [radius = 0.04];

    \draw [ultra thick, red] (-1, 0) arc (180 : 70 : 1);
  \end{tikzpicture}
  \caption
  {
    Example of a circular arc (red) that perfectly fits the vertices (blue) of the source polyline (black). However, the vertex shown by the green arrow did not appear due to the limitation of the format (inability to store circular arcs). This is because it does not follow the angular step of the vertices before this vertex ($ 18 $ vertices on the circular arc have $ 5\degree $ separation between them) and, therefore, should not be part of a circular arc.
  }
  \label{fig:FutureWorkUndesirableResultForArcFittingFromVertices}
\end{figure*}

The compression algorithm finds the resultant polyline within the tolerance of the source polyline and, therefore, can represent some part of the polyline with circular arcs where they were not present originally (see for example Fig.~\ref{fig:FutureWorkUndesirableResultForArcFittingFromVertices}). The decision of whether the part of the polyline can or cannot be represented as an arc has to be improved.

When the vertices of the resultant polyline are not required to be a subset of the vertices of the source polyline \cite{OptimalCompression}, higher compression can be achieved. Because this approach will likely improve performance and extend the algorithm described in \cite{OptimalCompression} to compression with arcs, it is a subject to future research.

\section*{Acknowledgment}

The author would like to thank Linda Thomas and Mary Anne Chan for proofreading this paper.


\newcommand{\doi}[1]{\textsc{doi}: \href{http://dx.doi.org/#1}{\nolinkurl{#1}}}


\bibliographystyle{IEEEtran}
\bibliography{AnEfficientCombinatorialAlgorithmForOptimalCompressionOfAPolylineWithSegmentsAndArcs}


\end{document}